\documentstyle[aps,preprint,epsfig]{revtex}

\begin{document}
\title{A Variational Approach in the Dissipative Nonlinear Schr\"{o}dinger Equation}
\author{Dagoberto S Freitas\footnote{dfreitas@uefs.br}}
\address{Departamento de F\'{\i}sica, Universidade Estadual de Feira de Santana,\\
44031-460, Feira de Santana, BA, Brazil.}
\author{Jairo R de Oliveira\footnote{jrocha@lftc.ufpe.br}}
\address{Departamento de F\'{\i}sica e Matem\'{a}tica, Universidade Federal Rural de\\
Pernambuco, 52171-030, Recife, PE, Brazil and Departamento de F\'{\i}sica,\\
Universidade Federal de Pernambuco, 50670-910, Recife, PE, Brazil.}
\date{\today }
\maketitle

\begin{abstract}
The properties of pulse propagation in a nonlinear fiber including linear
damped term added in the usual nonlinear Schr\"odinger equation is analyzed
analytically. We apply variational modified approach based on the lagrangian
that describe the dynamic of system and with a trial function we obtain a
solution which is more accuracy when compared with a pertubative solution.
As a result, the problem of pulse propagation in a fiber with loss can be
described in good agreement with exact results.
\end{abstract}

In recent years the propagation of optical pulses in fibers has obtained a
great attention not only from the theoretical as well as from the
experimental point of view. The nonlinear Schr\"odinger equation (NLSE) has
been employed to explain a variety of effects in the propagation of optical
pulses. As is well known the balance between the self-phase modulation (SPM)
and group velocity dispersion (GVD) leads to the so called solitons
solutions for the NLSE\cite{agrawal,zakharov}. Solitary wave
solutions have been known to exist in a variety of nonlinear and dispersive
media for many years. In the context of optical communications, Hasegawa and
Tappent \cite{hasegawa} first made the important observation that a pulse
propagating in an optical fiber with Kerr-law nonlinearity can form an
envelope soliton. This offered the potential for undistorted pulse
transmission over very long distances. Just as a balance between
self-phase-modulation and group-velocity dispersion can lead to the
formation of temporal solitons in single-mode fibers, it is also possible to
have the analogous spatial soliton, where diffraction and self-focusing can
compensate for each other \cite{zakharov}. The importance of studying
optical solitons is from the fact that their have potential applications in
optical transmission and all-optical processing. A soliton is a particular
solution of the nonlinear-wave equation. Since analytical solution are known for only
a few cases, investigations into the properties solutions are normally
performed numerically using such approaches. However, it is often desirable
to have an analytical model describing the dynamics of pulse propagation in
a fiber.

In the theoretical treatment of these problems, considerable attention has been
given to the variational approach \cite{firth,kalson}. A variational
approach was employed in \cite{kalson} deriving information about the
various parameters that characterize the beam, which are qualitatively as
well as quantitatively, in good agreement with numerical results. This
result invalidates the possibility of pulse compression without external
gratings which is erroneous and is only an artifact of the paraxial
approximation. In the same sense Anderson \cite
{anderson2} described the main characteristics of the temporal
soliton as determined by NLSE. The discussion above does not consider the
presence of the loss in the medium. It is well known that in real materials,
the medium will not be purely transparent and the nonlinearity will not be
of pure Kerr-law form, but will saturate. The problem of describing the
physical properties of dissipative systems has been the subject of lengthily
discussions \cite{ray,herrera}. These results were recently applied
to the problem of propagation of cw (continuous wave) Gaussian beams in a
saturable medium with loss \cite{jovanoski}. In that work \cite{jovanoski}
the diffraction is limited to one transverse solution. After that this problem
was analyzed using a variational modified approach \cite{dago}.

In this paper, we will analyses the dynamics interplay between nonlinearity
and dispersion through optical medium with loss using a variational modified
approach \cite{dago}. Exact analytical expressions for the behavior of the pulse
are determined.

The starting point of our analysis is the Nonlinear Schr\"{o}dinger Equation
that describe the propagation of a pulse envelope in a nonlinear loss medium,
\begin{equation}
i\frac{\partial u}{\partial \xi }+\frac{1}{2}\frac{\partial ^{2}u}{\partial
\tau ^{2}}+u\left| u\right| ^{2}=-i\Gamma u  \label{nls}
\end{equation}
where $u(\zeta ,\tau )$ is the normalized amplitude of the pulse, $\xi $
is the normalized coordinate, $\tau $ is the normalized time, and $\Gamma $
is the normalized loss parameter of the medium.

Now we can handle Eq.(\ref{nls}) adequately in the form,
\begin{equation}
\frac \partial {\partial \tau }\frac \partial {\partial u_\tau ^{*}}\left(
e^{\Gamma \xi }L\right) +\frac \partial {\partial \xi }\frac \partial {%
\partial u_\xi ^{*}}\left( e^{\Gamma \xi }L\right) -\frac \partial {\partial
u^{*}}\left( e^{\Gamma \xi }L\right) =0  \label{elmod}
\end{equation}
where
\[
L=\frac 12|\frac{\partial u}{\partial t}|^2+i(u\frac{\partial u^{*}}{%
\partial \xi }-u^{*}\frac{\partial u}{\partial \xi })-\frac 12|u|^4
\]
and $u^{*}$ is complex conjugate of $u$ and subindexes are the
differentiation with respect to $\tau $ and $\xi $ . $L$ is the lagrangian
of system without loss. The Eq.(\ref{elmod}) is the Euler-Lagrange equation
in the modified form that describe the propagation of the pulse in the
medium with loss, and can be written in the form of the modified Hamilton's
principle \cite{herrera,dago},
\begin{equation}
\delta \int_0^\infty \int_0^\infty e^{\Gamma \xi }Ld\xi d\tau =0
\label{phmod}
\end{equation}
Assuming a trial functional of the form
\begin{equation}
u\left( \xi ,\tau \right) =A\left( \xi \right)
\mathop{\rm sech}
\left( \frac \tau {w\left( \xi \right) }\right) \exp \left( i\phi \left( \xi
\right) \right) \text{,}  \label{sphmod}
\end{equation}
where $A$ is the amplitude of the pulse propagated, $w$ is the width and $\phi $
phase term. Using Eq.(\ref{sphmod}) into the variational formulation, Eq.(%
\ref{phmod}), we can integrate the $\tau $ dependence explicit to obtain
\begin{equation}
\delta \int_0^\infty e^{\Gamma \xi }\left\langle L\right\rangle d\xi=0,
\end{equation}
where
\begin{equation}
\left\langle L\right\rangle =\frac{\left| A\right| ^2}{3w}+2iw\left( A\frac{%
dA^{*}}{d\xi }-A^{*}\frac{dA}{d\xi }\right) +4w\left| A\right| ^2\frac{d\phi
}{d\xi }-2w\frac{\left| A\right| ^4}3
\end{equation}
is the average of $L$ in the time.
Then, from the standard calculus, deriving $e^{\Gamma \xi }\left\langle
L\right\rangle $ with respect to $A$, $A^{*}$, $w$ and $\phi $ we obtain the
following system of coupled ordinary differential equations
\begin{equation}
\frac d{d\xi }\left( w\left| A\right| ^2\right) =-\Gamma w\left| A\right| ^2
\label{pdiss1}
\end{equation}
\begin{equation}
w^2\left| A\right| ^2=1  \label{eqwA}
\end{equation}
\begin{equation}
8w\left| A\right| ^2\frac{d\phi }{d\xi }=\frac{8w\left| A\right| ^4}3-\frac{%
2\left| A\right| ^2}{3w}-4iw\left( A\frac{dA^{*}}{d\xi }-A^{*}\frac{dA}{d\xi
}\right) \text{.}  \label{fase}
\end{equation}
The equations above describe the characteristics of the pulse and solving
these equations we will obtain the full dynamics of the pulse through the
medium. It is obvious that once Eq.(\ref{pdiss1}) and Eq.(\ref{eqwA}) are
solved for $w$ and $\left| A\right| ^2$, the phase $\phi $ is easily
obtained from Eq.(\ref{fase}). In particular, if the longitudinal phase of
the amplitude $A$ is introduced by writing $A=\left| A\right| e^{i\theta
\left( \xi \right) }$ the Eq.(\ref{fase}) can be written as
\begin{equation}
\frac d{d\xi }\left( \phi +\theta \right) =\frac 1{4w^2}.
\end{equation}
from it we obtain
\begin{equation}
\phi \left( \xi \right) +\theta \left( \xi \right) =\frac 1{8\Gamma w\left(
0\right) ^2}\left( 1-e^{-2\Gamma \xi }\right) \text{.}  \label{rgfase}
\end{equation}
The equation above describe the regularized phase of the pulse. This system
of equation has analytic solution. From Eq.(\ref{pdiss1}) we obtain
\begin{equation}
w\left( \xi \right) \left| A\left( \xi \right) \right| ^2=w\left( 0\right)
\left| A\left( 0\right) \right| ^2e^{-\Gamma \xi }\text{.}  \label{pdiss2}
\end{equation}
The compatibility of Eqs.(\ref{pdiss2}) and (\ref{eqwA}) is possible when
\begin{equation}
\left| A\left( \xi \right) \right| =\left| A\left( 0\right) \right|
e^{-\Gamma \xi }
\end{equation}
and
\begin{equation}
w\left( \xi \right) =w\left( 0\right) e^{\Gamma \xi }\text{,}  \label{larg2}
\end{equation}
where was used the relation $\left| A\left( 0\right) \right| ^2=1/w^2\left(
0\right) $, with $A\left( 0\right) $ and $\omega \left( 0\right) $ is the
initial amplitude and width of pulse, respectively . Now we can write the
amplitude
\begin{equation}
A\left( \xi \right) =\left| A\left( 0\right) \right| e^{-\Gamma \xi
}e^{i\theta \left( \xi \right) }\text{.}
\end{equation}
Using the result above into the trial functional, Eq.(\ref{sphmod}), we can\
write the pulse in form
\begin{equation}
u\left( \xi ,\tau \right) =\left| A\left( 0\right) \right| e^{-\Gamma \xi
}\sec h\left( \frac \tau {w\left( \xi \right) }\right) \exp \left[ i\left(
\phi \left( \xi \right) +\theta \left( \xi \right) \right) \right] \text{,}
\label{intens2}
\end{equation}
where the regularized phase $\phi \left( \xi \right) +\theta \left( \xi
\right) $ is given by Eq.(\ref{rgfase}) \ and width $w\left( \xi \right) $
by Eq.(\ref{larg2}).

As would expect, the fiber loss is detrimental simply because the peak power
decreases exponentially with the fiber length. As a result, the pulse width
of the fundamental soliton also increase with propagation, as seen in the
figure. However, these results are qualitatively better than the results
obtained by using the inverse scattering method where $\Gamma $ is treated
as a weak pertubation. The our results foresee that the amplitude as well as
the width of the pulse suffer a smaller effect of the fiber loss that
thought, and approximate more of exact numerical solution by a factor of 2
in the exponent of the exponentials\cite{hasegawa,satsuma,agrawal2}.

In conclusion, the propagation of a pulse in a nonlinear loss medium has
been analysed using a variational modified approach. This modified approach
describes in a more consistent way the behavior of pulse in a dissipative
system. The our results are more accuracy when compared with a pertubative
solution where $\Gamma $ is treated as a weak pertubation.

\acknowledgements{One of us (J.R.O) thanks the financial support by
Conselho Nacional de Desenvolvimento Cient\'{\i}fico e Tecnol\'ogico
(CNPq), Brazil.}

\newpage

\begin{figure}[tbp]
\vspace{1cm}
\begin{center}
\epsfig{file=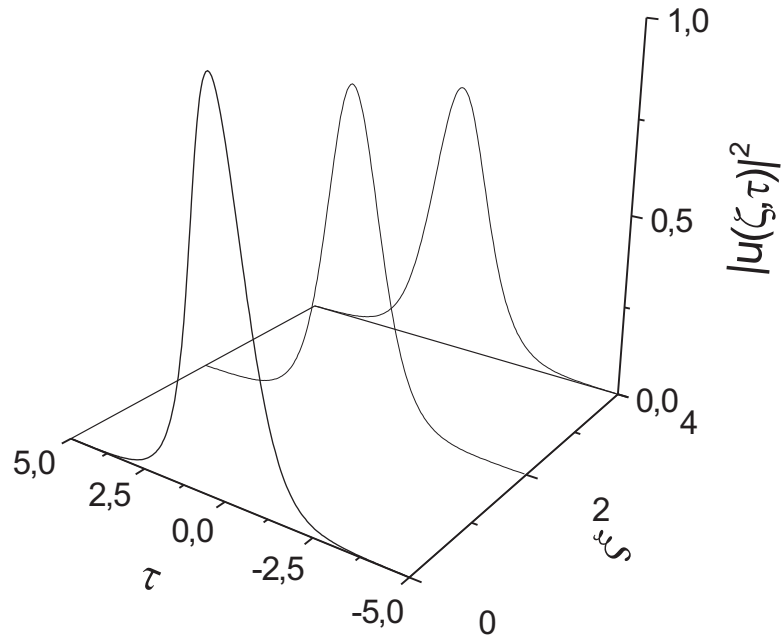,height=4.0truein}
\end{center}
\vspace{1cm}
\protect\caption{Plot illustrate the pulse propagation for $\xi=0,~2$ and
$4$, taking $\Gamma=0.035$ }
\label{fig}
\end{figure}


\begin{references}

\bibitem{agrawal}  G. P. Agrawal, Nonlinear Fiber Optics (Academic, San
Diego, 1989)

\bibitem{zakharov}  V. E. Zakharov and A. B. Shabat, Zh. Eksp. Teor. Fiz.
61, 118(1971) [Sov. Phys. JETP 34, 62 (1972)].

\bibitem{hasegawa}  A. Hasegawa and F. Tappert, Appl. Phys. Lett. 23, 142
(1973).

\bibitem{firth}  W. J. Firth, Opt. Commun. 22, 226 (1977).

\bibitem{anderson1}  D. Anderson, M. Bonnedal and M. Lisak, Phys. Fluids.
22, 1838 (1979).

\bibitem{kalson}  M. Karlsson, D. Anderson, M. Desaix and M. Lisak, Opt.
Lett. 16, 1973 (1991).

\bibitem{anderson2}  D. Anderson, Phys. Rev. A27, 3135 (1983).

\bibitem{chiao}  R. Y. Chiao, E. Garmire, and C. H. Yownes, Phys. Rev. Lett.
13, 479 (1964).

\bibitem{max}  C. E. Max, Phys. Fluids. 19, 74 (1976).

\bibitem{sodha}  M. S. Sodha and V. K. Tripathi, Phys. Rev. A16, 201 (1977).

\bibitem{manassah1}  J. T. Manassah, P. L. Baldeck and R. R. Alfano, Opt.
Lett. 13, 1090 (1988).

\bibitem{manassah2}  J. T. Manassah, P. L. Baldeck and R. R. Alfano, Opt.
Lett. 13, 589 (1988).

\bibitem{jovanoski}  Z. Jovanoski and R. A. Sammut, Phys. Rev. E50, 4087
(1994).

\bibitem{ray}  J. R. Ray, Am. J. Phys. 47, 626 (1979).

\bibitem{herrera}  L. Herrera, L. N\'u\~nez, A. Pati\~no and H. Rago, Am. J.
Phys. 54, 273 (1986).

\bibitem{dago}  D. S. Freitas, J. R. de Oliveira and M. A. de Moura, J.
Phys. A: Math. Gen. 30 (1997).

\bibitem{kogelnik}  H. Kogelnik, T. Li, Appl. Opt.5, 1550 (1966).

\bibitem{satsuma}  J. Satsuma and N. Yajiama, Prog. Theor. Phys. Suppl. 55,
284 (1974)

\bibitem{agrawal2}  G. P. Agrawal, Nonlinear Fiber Optics (Academic, San
Diego, 1989)[see ch. 5.4]
\end{references}
\end{document}